\def\ln{\ell{n}}
\documentstyle[12pt]{article}
\begin{document}
\begin{titlepage} \vspace{0.2in} \begin{flushright}
MITH-97/3 \\ \end{flushright} \vspace*{0.5cm}
\begin{center} {\LARGE \bf  Fine-tuning the low-energy physics\\} 
\vspace*{0.8cm}
{\bf She-Sheng Xue}\\ \vspace*{1cm}
Dipartimento di Fisica, Universit\'a di Milano\\
INFN - Sezione di Milano, Via Celoria 16, Milan, Italy\\
\vspace*{1.5cm}
{\bf   Abstract  \\ } \end{center} \indent

In this article we discuss the problem of the extreme fine-tuning necessary to
achieve the top quark mass scale $m_t$ within the dynamical symmetry breaking of
the phenomenological $t \bar t$ condensate model. Inspired by the vector-like
phenomenon of chiral gauge theories at short distances, we postulate that the
$W^\pm$-gauge bosons possess a vector-like gauge coupling and contribute to the
gap-equations of quark self-energy functions in the high-energy region. We find
that the gap-equations of different charge sectors are coupled, and as a 
result the unnatural fine-tuning is overcome and the ratio of the top and
bottom masses is related to the electric charge: $\left({m_b\over
m_t}\right)^2\simeq {\alpha\over3\pi}$, provided the quadratic divergence is
removed by setting the four-fermion coupling $G=4+o\left({m_b^2\over
m_t^2}\right)$. 

\vfill \begin{flushleft}  December, 1996   \vspace*{2cm} \\
\noindent{\rule[-.3cm]{5cm}{.02cm}} \\
\vspace*{0.2cm} \hspace*{0.5cm} ${}^{a)}$ 
E-mail address: xue@milano.infn.it\end{flushleft} \end{titlepage}

\noindent
{\bf 1.~Introduction}

It is well known that the dynamical symmetry breaking of the Nambu-Jona Lasinio
(NJL) type \cite{njl} meets the severe problem of extreme fine-tuning. Indeed
the gap equation obeyed by the mass gap turns out to be quadratically divergent
($O(\Lambda^2)$), and in the absence of a symmetry that protects the mass gap
from such divergence, the only way out is to tune the NJL coupling to a
critical value with the incredible precision of $\left({m\over
\Lambda}\right)^2$. And this is clearly very unnatural. Analogously, the vacuum
expectation value of the elementary scalar field envisaged in the Higgs
mechanism is also plagued by quadratic divergences, stemming from radiative
corrections, and a similar fine-tuning of the $\phi^4$-potential is needed in
order to eliminate the quadratic divergences. Supersymmetry with its implied
cancellation of the quadratic divergences between the bosonic and fermionic
sectors of the theory thus becomes a very strong theoretical candidate to cure
such problem. This suggests that the solution of the problem implies a highly
non-trivial structure of the theory in the high-energy region. 

In this article we wish to show that there is another solution,
which is much more economical in terms of particles and interactions. In
section 2, we focus on the third quark family to carefully describe 
fine-tuning in the context of dynamical symmetry breaking of the
phenomenological $t \bar t$ condensate model. In section 3, we discuss our
motivation and hypothesis of the $W^\pm$ gauge boson's coupling and its
contribution to the Schwinger-Dyson equations of fermion masses. Such coupled 
Schwinger-Dyson
equations are solved in section 4. The discussions of how the problem of
fine-tuning is evaded and some remarks are presented in the last section. 

\vskip0.3cm
\noindent
{\bf 2.~The problem of fine-tuning}

In order to explicitly show how we obtain the low-energy weak
scale by dynamical symmetry breakings, we consider the top-condensate 
BHL-model\cite{bar} for the third quark generation
of the top and bottom quarks. This model 
is phenomenologically\footnote{In ref.\cite{xue94}, we have shown
that the solution of only one very massive quark is energetically
favourable.} given by the lagrangian,
\begin{equation}
L=L_{\rm standard}
+G_\circ\bar \Psi^{ia}_L(x)\cdot t_R^a(x)\bar t_R^b(x)\cdot \Psi^{ib}_L(x),
\label{tt}
\end{equation}
where $\Psi^i_L(x)$ is the left-handed weak doublet and $a,b$ are color indices
and $G_\circ$ is the four-fermion coupling, due to some underlying physics at the 
cutoff scale $\Lambda$. When the four-fermion coupling and the electric 
coupling (charge) are
larger than certain critical values, this model develops the phenomenon of
spontaneous symmetry breaking and a non-zero fermion self-energy function
$\Sigma_t(p)$ of the top quark emerges. To illustrate 
fine-tuning needed to obtain $\Sigma_t(p)\ll\Lambda$ in the low-energy limit, 
first we briefly 
summarize some basic features about the Schwinger-Dyson approach 
that were discussed in the refs.\cite{kogut}.

Owing to its purely left-handed gauge coupling to fermions, the $W$-boson
does not contribute to the Schwinger-Dyson equations. Note that we ignore the 
$Z$-boson's
contribution for its perturbative character and the QCD-contribution for its confining
property. We only take the QED and the four-fermion interaction into account.
In the Landau gauge, the fermion self-energy functions $\Sigma_{t,b}(p)$ for 
the top and bottom quark satisfy,
\begin{eqnarray}
\Sigma_t(p^2)&=&m-{G_\circ\over2}\langle\bar t t \rangle +
3Q^2\int^\Lambda_{p'} {1\over (p-p')^2}
{\Sigma_t(p'^2)\over p'^2+\Sigma_t^2(p'^2)};
\label{self}\\
\Sigma_b(p^2)&=&m+3{Q^2\over4}\int^\Lambda_{p'} {1\over (p-p')^2}
{\Sigma_b(p'^2)\over p'^2+\Sigma_b^2(p'^2)},
\label{self'}
\end{eqnarray}
where $m$ is the bare mass added to the lagrangian (\ref{tt}), 
$Q={2\over3}e$ is the bare charge of the top quark at the cutoff. 
After performing the angular 
integration and changing variables to $x=p^2$, eqs.(\ref{self},\ref{self'}) 
can be converted to the following boundary value problems:
\begin{eqnarray}
{d\over dx}\left(x^2\Sigma_t'(x)\right)+{\alpha_t\over 4\alpha_c}{x\over x+
\Sigma_t^2(x)}\Sigma_t(x)&=&0,\label{deq}\\
(1+g)\Lambda^2\Sigma_t'(\Lambda^2)+\Sigma_t(\Lambda^2)&=&m,
\label{boundary}
\end{eqnarray}
and
\begin{eqnarray}
{d\over dx}\left(x^2\Sigma_b'(x)\right)+{\alpha_t\over 16\alpha_c}{x\over x+
\Sigma^2_b(x)}\Sigma_b(x)&=&0,\label{deqb}\\
\Lambda^2\Sigma_b'(\Lambda^2)+\Sigma_b(\Lambda^2)&=&m,
\label{boundaryb}
\end{eqnarray}
where the notations are:
\begin{equation}
\alpha_t={Q^2(\Lambda)\over 4\pi};\hskip0.3cm\alpha_c={\pi\over 3};\hskip0.3cm 
g={N_cG_\circ\Lambda^2\over2\pi^2}{\alpha_c\over\alpha_t};\hskip0.3cm
\langle\bar tt\rangle=\Lambda^4{\Sigma_t'(\Lambda^2)\over 3\pi\alpha_t}.
\label{defintions}
\end{equation}
Note that the bare mass $m$ and the four-fermion coupling $g$ enter the gap 
equation
through the boundary condition only. For $x\gg 1$, the nonlinearity present 
in the boundary value problem can be neglected,  
the solutions for the self-energy functions in the ultraviolet region are 
given,
\begin{eqnarray}
\Sigma_t(x) &=& {A_t\mu^2\over\sqrt{ x}}{\rm sinh}\left({1\over2}\sqrt{
1-{\alpha_t\over\alpha_c}}\ln({x\over\mu^2})\right);
\label{solution}\\
\Sigma_b(x)&=&{A_b\mu^2\over\sqrt{x}}{\rm sinh}\left({1\over2}\sqrt{
1-{\alpha_t\over4\alpha_c}}\ln({x\over \mu^2})\right),
\label{solutionb}
\end{eqnarray}
where $A_{t,b}$ are finite constants, that should be related to the Yukawa
couplings of the top and bottom quarks, and $\mu$ some infrared scale. The
$A$'s and $\mu$, in principle, are determined by the boundary conditions at
both the infrared limit and the ultraviolet limit. Note that
solutions (\ref{solution}) and (\ref{solutionb}) hold in the
ultraviolet limit, thus we are only allowed to enforce the ultraviolet boundary
conditions (\ref{boundary},\ref{boundaryb}). Using these boundary conditions, 
we obtain the gap-equations: 
\begin{eqnarray}
m&=&{A_t\mu^2\over2\Lambda}\left[(1-g){\rm sinh}\theta+
(1+g)\sqrt{1-{\alpha_t\over\alpha_c}} {\rm cosh}\theta\right];
\label{boundary1}\\
m&=&{A_b\mu^2\over2\Lambda}\left[{\rm sinh}\theta'+
\sqrt{1-{\alpha_t\over4\alpha_c}} {\rm cosh}\theta'\right],
\label{boundary1b}
\end{eqnarray}
where 
\begin{equation}
\theta={1\over2}\sqrt{
1-{\alpha_t\over\alpha_c}}\ln({\Lambda^2\over\mu^2}),
\label{theta}
\end{equation}
and $\theta'$ is defined as eq.(\ref{theta}) with substitution 
$\alpha_t\rightarrow {\alpha_t\over4}$.
The gap-equations (\ref{boundary1},\ref{boundary1b}) for the top and bottom
quarks can be expressed as 
\begin{eqnarray}
m&=&{A_t\mu\over4}\left({\mu\over\Lambda}\right)^{
1-\sqrt{1-{\alpha_t\over\alpha_c}}}
\left[1-g+(1+g)\sqrt{1-{\alpha_t\over\alpha_c}}\right]\nonumber\\
&+&{A_t\mu\over4}\left({\mu\over\Lambda}\right)^{
1+\sqrt{1-{\alpha_t\over\alpha_c}}}
\left[(1+g)\sqrt{1-{\alpha_t\over\alpha_c}}-(1-g)\right].
\label{low3}\\
m&=&{A_b\mu\over4}\left({\mu\over\Lambda}\right)^{
1-\sqrt{1-{\alpha_t\over4\alpha_c}}}
\left[1+\sqrt{1-{\alpha_t\over4\alpha_c}}\right]\nonumber\\
&+&{A_b\mu\over4}\left({\mu\over\Lambda}\right)^{
1+\sqrt{1-{\alpha_t\over4\alpha_c}}}
\left[\sqrt{1-{\alpha_t\over4\alpha_c}}-1\right].
\label{low3b}
\end{eqnarray}
These are the gap-equations that minimize the vacuum energy of the theory. Given
$g$ and $\alpha_t$, these equations establish the relationship between the cutoff 
$\Lambda$, the infrared limit $\mu$ and the $A$'s. 

In the case of explicit symmetry breaking ($m\not=0$), we have 
($\mu\ll\Lambda$)
\begin{eqnarray}
m&=&{A_t\mu\over4}\left({\mu\over\Lambda}\right)^{\alpha_t\over2\alpha_c}
\left[2-(1+g){\alpha_t\over2\alpha_c}\right],
\label{low33}\\
m&=&{A_b\mu\over4}\left({\mu\over\Lambda}\right)^{
\alpha_t\over8\alpha_c}
\left[2-{\alpha_t\over8\alpha_c}\right].
\label{low33b}
\end{eqnarray}
The top quark mass $m_t=A_t\mu$ and the bottom quark mass $m_b=A_b\mu$
are related to the finite bare mass $m$, their scaling behaviours in the
infrared limit (${\mu\over\Lambda}\ll 1$) are given by the equations
(\ref{low33},\ref{low33b}). No fine-tuning of the couplings $g$ and $\alpha_t$
is needed. 

In the chiral limit ($m=0$), i.e.~when no explicit symmetry breaking occurs, 
we study how the low-energy limit ($\mu\ll\Lambda$) is obtained
after the spontaneous symmetry breaking. 
From the gap-equation (\ref{low3}) one finds for large values of the 
four-fermion coupling $g$ and QED coupling $\alpha_t$,
\begin{equation}
A_t\not=0,\hskip0.3cm\left({\mu\over\Lambda}\right)^2=\left[{1-g+(1+g)
\sqrt{1-{\alpha_t\over\alpha_c}}\over 2(1-g)}
\right]^{1\over
\sqrt{1-{\alpha_t\over\alpha_c}}},
\label{low}
\end{equation}
showing that the infrared scale $\mu^2$ is directly proportional to the 
ultraviolet scale $\Lambda^2$, which is the unique scale in the theory. As a
result the
top quark mass $m_t\sim A_t\mu$ is generated by the spontaneous symmetry breaking.
Due to the absence of the four-fermion coupling in the gap-equation
(\ref{low3b}) for the bottom quark, there exists only the trivial solution, 
i.e., the
self-energy function $\Sigma_b(p)$ of the bottom quark 
in the chiral limit and small QED gauge coupling is identically
zero, 
\begin{equation}
A_b=0\hskip0.5cm\Sigma_b(p)\equiv 0.
\label{b0}
\end{equation}
The bottom quark remains
massless because its Schwinger-Dyson equation is homogeneous\cite{mn}. 

Once the system undergoes the spontaneous symmetry breaking, 
there are Goldstone bosons\cite{njl,g} above infinitely degenerate ground states and
the value
of the infrared scale $\mu$ totally depends on the couplings
$g$ and $\alpha_t$ chosen. All ``Yukawa coupling'' $A$'s are independent 
parameters that do not enter the gap equations. 
In order to achieve the right weak scale ($\mu\sim 250$GeV), we need an 
extreme fine-tuning of the four-fermion coupling $G$,
\begin{equation}
\left({\mu\over\Lambda}\right)^2={\alpha_t\over4\alpha_c}{ G+{\alpha_t\over
\alpha_c}-4\over G-{\alpha_t\over\alpha_c}}
\sim 10^{-34};\hskip0.3cm G\equiv{N_cG_\circ\Lambda^2\over2\pi^2},
\label{fine}
\end{equation}
where the cutoff is assumed at the Planck scale and the
gauge coupling ${\alpha_t\over\alpha_c}\ll 1$. The critical value is thus
given by $ G+{\alpha_t\over
\alpha_c}=4$. As already noted, this fine-tuning is due to the existence of 
the quadratic
divergence in the gap-equation (\ref{fine}) for the self-energy 
function. 
The fact that a totally unnatural fine-tuning is forced in
order to not only obtain non-perturbative dynamical symmetry breaking $\mu\not=
0$, but also remove the quadratic divergence to reach the low-energy 
limit $\mu\ll\Lambda$ signals a
non-trivial structure of the ground states in the high-energy region. 

\vskip0.5cm
\noindent
{\bf 2.~The vector-like phenomenon and three-fermion states}
\vskip0.3cm

Due to the underlying physics at the cutoff, beside the dimension-6
operator (\ref{tt}), the effective lagrangian could contain other high 
dimensional
operators, which obey the gauge symmetries of the Standard Model. The 
effective coupling
of these operators could be momentum-dependent. When the
effective couplings are larger than a certain threshold value $\epsilon$,
three-fermion Weyl states and other composite states with appropriate 
quantum numbers are bound, and the mass spectra and the 1PI vertices are 
vector-like,
consistently with chiral gauge symmetries of the Standard Model. We will not enter into the details
of this phenomenon that is described in ref.\cite{xue97}. In this article, instead, we wish to reconsider the Schwinger-Dyson
equations (\ref{self},\ref{self'}) for the self-energy functions $\Sigma_t(p)$ 
and
$\Sigma_b(p)$ of the top and bottom quarks by taking into account the possible
relevant 1PI vertex functions arising in the high-energy region. Thus, inspired
by such vector-like phenomenon in the high energy region, we postulate  
beyond a certain energy scale $\epsilon$, larger than the weak scale, 

\begin{enumerate}
\begin{itemize}

\item  the existence of right-handed three-fermion Weyl states with 
the chiral gauge quantum
numbers of the weak $SU_L(2)$ group. These states 
combine with elementary Weyl
fermions to form massive Dirac fermions;

\item the vector-like character of the coupling vertex between
these composite Dirac fermions and the $W^\pm$ gauge boson. 

\end{itemize}
\end{enumerate}

We are now going to explicitly discuss these two 
assumptions within the third quark family in the high energy region. We assume there is an 
intermediate
energy-threshold $\epsilon$ between the weak scale ($\mu\sim 250$GeV) and the
cutoff $\Lambda$, 
\begin{equation}
250GeV <\epsilon < \Lambda,
\label{epsilon}
\end{equation}
above which the effective couplings of high dimensional operators are strong 
enough to form the three-fermion bound states of the third quark family that 
are given by,
\begin{equation}
T_R\sim (\bar t_R\cdot t_L)t_R,\hskip0.3cm B_R\sim (\bar b_R\cdot b_L)b_R,
\label{threeb}
\end{equation}
which are right-handed Weyl fermions with the appropriate gauge quantum number
of $SU_L(2)$. On the other hand, as the effective couplings decrease, these right-handed
three-fermion bound states disappear (turning to unbound three fermion states
) at the threshold $\epsilon$ (\ref{epsilon}), to reproduce the
parity-violating gauge coupling of the low-energy phenomenology, 

In the high energy region, the right-handed fermion state comprises
the elementary state $t_R$ ($b_R$) and the composite state $T_R$ ($B_R$): 
\begin{equation}
\Psi^t_R=\{t_R,T_R\};\hskip0.5cm \Psi^b_R=\{b_R,B_R\}.
\label{mixing}
\end{equation}
The composite Dirac particles are then given,
\begin{equation}
\Psi^t_D=\{t_L,\Psi^t_R\};\hskip0.5cm \Psi^b_D=\{b_L,\Psi^b_R\},
\label{mixingd}
\end{equation}
which couple to the $W^\pm$-gauge boson in a vector-like manner.
Thus the effective gauge coupling of $W^\pm$-bosons to
quarks in the high-energy region becomes vector-like. For the quark-$W^\pm$
vertex one may write: 
\begin{eqnarray}
\Gamma^w_\mu (q)&=&ig_w(q)\gamma_\mu (P_L+f(q))\label{wv}\\
f(q)&\not=&0\hskip0.2cm {\rm for}\hskip0.2cm q >\epsilon,
\end{eqnarray}
where the CKM-matrix\cite{ckm} is set to be identity and $g_w(q)$ is a renormalized coupling constant.
Below the energy threshold (\ref{epsilon}) where the three-fermion states 
dissolve into their constituents, we assume that the
vertex function $f(q)$ vanishes, i.e.
\begin{equation}
f(q)|_{q\rightarrow\epsilon}=0.
\label{threshold}
\end{equation}
The scenario of the vector-like phenomenon above the intermediate scale (\ref{epsilon}) is
reminiscent of ``left-right'' symmetric extensions of the Standard
model\cite{lr}. 

The self-energy functions $\Sigma_{t,b}(p)$ in the high-energy region ($p\gg
1$) should be defined as the truncated two-point functions of the elementary
left-handed fields ($t_L,b_L$) and the right-handed mixing states ($\Psi_R^t,
\Psi_R^b$) (\ref{mixing}). Thus, we find that the $W^\pm$ bosons give the following
contributions to the RHS of the Schwinger-Dyson equations
(\ref{self},\ref{self'}) in the high-energy region $(p'\ge\epsilon)$: 
\begin{eqnarray}
W_b(p)&=&\int^\Lambda_{|p'|\ge\epsilon} {Q_w^2\over (p-p')^2+M_w^2}
{\Sigma_t(p'^2)\over p'^2+\Sigma_t^2(p'^2)};\nonumber\\
W_t(p)&=&\int^\Lambda_{|p'|\ge\epsilon} {Q_w^2\over (p-p')^2+M_w^2}
{\Sigma_b(p'^2)\over p'^2+\Sigma_b^2(p'^2)},
\label{w}
\end{eqnarray}
where $Q_w$ is the weak charge (\ref{wv}), and the integration of the internal  momentum $p'$
starts from the intermediate threshold $\epsilon$ to the cut-off $\Lambda$.

Due to the properties of the right-handed composite states $T_R$ and $B_R$
(\ref{threeb}) that carry the appropriate quantum numbers and couple to the
$W^\pm$-gauge bosons, the fermionic self-energy functions $\Sigma_{t,b}(p)$ do not
violate the chiral gauge symmetries of the Standard model in the high-energy
region ($p>\epsilon$), where the spontaneous symmetry breaking is soft and thus
irrelevant. Thus the self-energy functions $\Sigma_{t,b}(p)(p>\epsilon)$ can be non-vanishing
without violating the $SU_L(2)$ chiral gauge symmetry, and the corresponding Ward
identity relates the self-energy functions  $\Sigma_{t,b}(p)$ to the
vector-like vertex function $f(q)$ (\ref{wv}). 

\vspace*{0.5cm}
\noindent
{\bf 3.~Solution to the coupled Schwinger-Dyson equations}
\vskip0.3cm

The $W$-bosons perturbatively contribute to the original
Schwinger-Dyson equations for the top quark (\ref{self}) and the bottom quark
(\ref{self'}) respectively. One can see that eq.(\ref{w}) mixes the
Schwinger-Dyson equations for the fermionic self-energy functions of different
charge sectors. As a consequence,
eqs.~(\ref{self},\ref{self'}) are no longer independent 
equations, instead, they are perturbatively coupled together:
\begin{eqnarray}
\Sigma_t(p^2)&=&-{G_\circ\over2}\langle\bar t t \rangle
+W_t(p)+3Q^2\int^\Lambda_{p'} {1\over (p-p')^2}
{\Sigma_t(p'^2)\over p'^2+\Sigma_t^2(p'^2)};\label{selft}\\
\Sigma_b(p^2)&=&W_b(p)+{3Q^2\over4}\int^\Lambda_{p'} {1\over (p-p')^2}
{\Sigma_b(p'^2)\over p'^2+\Sigma_b^2(p'^2)},\label{selfb}
\end{eqnarray}
where the bare masses $m$ are set to zero. For $p\gg 1$ and $|p'|\ge\epsilon
\gg 1$, we approximate the W's contributions (\ref{w}) to be,
\begin{equation}
W_{b,t}(p)=\int^\Lambda_{|p'|\ge\epsilon} {Q_w^2\over (p-p')^2+M_w^2}
{\Sigma_{t,b}(p'^2)\over p'^2+\Sigma_{t,b}^2(p'^2)}
\simeq\alpha_w\Sigma_{t,b}(\Lambda),
\label{w'}
\end{equation}
where $\alpha_w$ is an undetermined constant.

Analogously to eqs.(\ref{deq},\ref{deqb}),
in the ultraviolet region ($x=p^2\gg 1$), 
the integral eqs.(\ref{selft},\ref{selfb}) can be converted to the following 
boundary value problems:
\begin{eqnarray}
{d\over dx}\left(x^2\Sigma_t'(x)\right)+{\alpha_t\over 4\alpha_c}
\Sigma_t(x)&=&0,
\label{deqt'}\\
(1+g)\Lambda^2\Sigma_t'(\Lambda^2)+\Sigma_t(\Lambda^2)&=&\alpha_w
\Sigma_b(\Lambda),
\label{boundaryt'}
\end{eqnarray}
and 
\begin{eqnarray}
{d\over dx}\left(x^2\Sigma_b'(x)\right)+{\alpha_t\over 16\alpha_c}
\Sigma_b(x)&=&0,\label{deqb'}\\
\Lambda^2\Sigma_b'(\Lambda^2)+\Sigma_b(\Lambda^2)&=&\alpha_w\Sigma_t(\Lambda).
\label{boundaryb'}
\end{eqnarray}
These differential equations are the same as before, except for
the fact that now the ultraviolet boundary conditions are coupled. 
Thus, substituting the known generic solutions 
(\ref{solution},\ref{solutionb}) for $x\gg 1$
into the boundary conditions (\ref{boundaryt'}) and
(\ref{boundaryb'}), we obtain the self-consistent gap-equations for 
$\Sigma_t(\Lambda)$ and $\Sigma_b(\Lambda)$, 
\begin{eqnarray}
\alpha_w\Sigma_b(\Lambda)&=&
{A_t\mu^2\over2\Lambda}\left[(1-g){\rm sinh}\theta+
(1+g)\sqrt{1-{\alpha_t\over\alpha_c}} {\rm cosh}\theta\right],
\label{boundary1f}\\
\alpha_w\Sigma_t(\Lambda)&=&{A_b\mu^2\over2\Lambda}\left[{\rm sinh}\theta'+
\sqrt{1-{\alpha_t\over4\alpha_c}} {\rm cosh}\theta'\right],
\label{boundary2f}
\end{eqnarray}
which differ in an essential way from the self-consistent gap-equations 
(\ref{boundary1},\ref{boundary1b}) for the decoupled system.

The first conclusion that one can derive from these gap-equations is that
the self-energy function of the bottom quark must be non-trivial 
\begin{equation}
\Sigma_b(\Lambda)\not=0;\hskip0.3cm {\rm if} \hskip0.3cm 
\Sigma_t(\Lambda)\not=0,
\end{equation}
where the $\Sigma_t(\Lambda)$ is generated by the spontaneous symmetry breaking
for large value of the four-fermion coupling $G_\circ$. The bottom quark mass
is generated by the explicit symmetry breaking, due to the top quark mass.
There is another trivial-solution: 
\begin{equation}
\Sigma_b(\Lambda)=0;\hskip0.3cm {\rm and} \hskip0.3cm 
\Sigma_t(\Lambda)=0.
\end{equation}
These solutions are not energetically favorable for large coupling $G_\circ$.

We now turn to find the solution of the gap-equations 
(\ref{boundary1f},\ref{boundary2f}) 
in the low-energy limit ($\mu\ll\Lambda$). These equations become
\begin{eqnarray}
\alpha_w\Sigma_b(\Lambda)&=&{A_t\mu\over4}\left({\mu\over\Lambda}\right)^{
1-\sqrt{1-{\alpha_t\over\alpha_c}}}
\left[1-g+(1+g)\sqrt{1-{\alpha_t\over\alpha_c}}\right];\label{low3'}\\
\alpha_w\Sigma_t(\Lambda)&=&{A_b\mu\over4}\left({\mu\over\Lambda}\right)^{
1-\sqrt{1-{\alpha_t\over4\alpha_c}}}
\left[1+\sqrt{1-{\alpha_t\over4\alpha_c}}\right].
\label{low3b'}
\end{eqnarray}
Taking the ratio of these two equations, we obtain
\begin{equation}
{\Sigma_b(\Lambda)\over
\Sigma_t(\Lambda)}=\left({A_t\mu\over A_b\mu}\right)\left({\mu\over\Lambda}
\right)^{3\alpha_t\over8\alpha_c}\left[{
1-g+(1+g)\sqrt{1-{\alpha_t\over\alpha_c}}\over
1+\sqrt{1-{\alpha_t\over4\alpha_c}}}\right].
\label{r}
\end{equation}
Considering the effect of renormalization on the gauge interactions (QCD, QED) 
and fermionic mass operators, we obtain
\begin{equation}
{\Sigma_b(\Lambda)\over\Sigma_t(\Lambda)}=
{\Sigma_b(\mu)\over\Sigma_t(\mu)}
\left({\alpha_t(\Lambda)\over\alpha_t(\mu)}\right)^{9\over20N_F}.
\label{re}
\end{equation}
where $N_F$ is the number of quark flavours.
We can appropriately choose constants $A_t$ and $A_b$ so that the masses 
of the top and bottom quarks at the infrared scale $\mu$ are given by,
\begin{equation}
\Sigma_t(\mu)\simeq A_t\mu;\hskip0.5cm
\Sigma_b(\mu)\simeq -A_b\mu,
\label{mu}
\end{equation}
where we attribute the fermion masses $m_{t,b}$ an opposite sign.
Thus, we get the following relationship
\begin{equation}
\left({\Sigma_b(\mu)\over
\Sigma_t(\mu)}\right)^2\left({\alpha_t(\Lambda)\over\alpha_t(\mu)}\right)^{9\over20N_F}
\left(1+{3\alpha_t(\Lambda)\over8\alpha_c}\ln {\Lambda\over\mu}
\right)={G+{\alpha_t(\Lambda)\over\alpha_c}-4\over
4-{\alpha_t(\Lambda)\over4\alpha_c}}.
\label{rr}
\end{equation}
This is a gap equation relating the masses $\Sigma_{t,b}(\mu)$ of the top and 
bottom quarks and the 
couplings $G,\alpha_t$, similar to the gap-equations (\ref{low33},\ref{low33b})
obtained in the case of explicit symmetry breakings.

\vspace*{0.5cm}
\noindent
{\bf 4.~Discussions and remarks}
\vskip0.3cm

As remarked after eq.(\ref{fine}), the problem of unnatural fine-tuning 
arises from the fact 
that we must simultaneously arrange the couplings $G$ and $\alpha_t$ in such
a way that the the quadratic divergence in the cut-off $\Lambda$ gets 
cancelled and in addition  the soft spontaneous symmetry breaking is 
non-trivial yielding $\mu\not=0, \mu\ll\Lambda$. On the other hand,
the coupled gap-equations (\ref{low3'},\ref{low3b'}) have eq.(\ref{rr})
as their implication, showing that we can dispose of the the cancellation 
of the quadratic divergence $\Lambda^2$ and of the non-trivial soft
spontaneous symmetry breaking in an independent way. Indeed we may assume that
$G$ goes exactly to the fixed point, i.e.
\begin{equation}
G\rightarrow 4+0^+,
\label{limit}
\end{equation}
where the $\Lambda^2$ is exactly cancelled and the soft spontaneous symmetry 
breaking ($\mu\not=0$)
gets in touch with the cut-off $\Lambda$ only logarithmically through the
relationship:
\begin{equation}
\left({\Sigma_b(\mu)\over
\Sigma_t(\mu)}\right)^2\left({\alpha_t(\Lambda)\over\alpha_t(\mu)}\right)^{9\over20N_F}
\left(1+{3\alpha_t(\Lambda)\over8\alpha_c}\ln {\Lambda\over\mu}
\right)\simeq {{\alpha_t(\Lambda)\over\alpha_c}\over
4-{\alpha_t(\Lambda)\over4\alpha_c}},
\label{rrr}
\end{equation}
obtained from eq(\ref{rr}) through (\ref{limit}). In eq.(\ref{rrr}),
no fine-tuning of the electric coupling $\alpha$ is needed to have $\mu\ll\Lambda$.

The reason for this appearing result is that we are approaching
the physical fixed point, i.e.~$(G,\alpha_t)\rightarrow 
(4, {4\over9}{1\over137})$, from the spontaneous symmetry
broken phase. At this fixed point, the effective dimensions of the four-fermion
operator (\ref{tt}) becomes 4 \cite{kogut} and the $\Lambda$-dependence is only
logarithmic. Note that this is true for both the the decoupled gap-equation
(\ref{fine}) and coupled gap-equation (\ref{rr}). However in the decoupled case
we must fine-tune the approach to the fixed point up to an unreasonable
precision $O\left(\mu^2\over\Lambda^2\right)$, while in the coupled case
we must demand that $G\rightarrow 4$ only up to 
$o\left({m^2_b\over m_t^2}\right)$. If we are willing to accept that, for some
general principle still to be studied, $G=4+0^+$ (or else 
$G=4+o\left({m^2_b\over m_t^2}\right)$), we obtain an interesting prediction
on the mass ratio $\left({m_b\over m_t}\right)$. In fact, 
taking the value $\alpha^{-1}(M_Z)\simeq 128$\cite{data} and $\Lambda=10^{19}$GeV, 
by renormalization group equation we obtain 
$\alpha^{-1}(\Lambda)\simeq 103$. From
equation (\ref{rrr}), one can make the estimate 
\begin{equation}
\left({m_b(M_Z)\over
m_t(M_Z)}\right)^2\simeq 0.93 {\alpha_t\over3\pi}\hskip0.5cm \alpha_t={1\over103},
\label{res}
\end{equation}
which is in consistent with experiment values.

We wish to thank Profs.~Y.~Nambu, E.~Akhmedov, 
M.~Lindner, H.Q.~Zheng for stimulating discussions.

\end{document}